\documentclass[aps,prb,twocolumn,showpacs,amsmath,amssymb,superscriptaddress,floatfix]{revtex4-1}
\usepackage{mathtools}
\usepackage{graphicx}
\usepackage{epsfig}
\usepackage{color}
\usepackage{bbm}
\usepackage{mathtools}
\usepackage{hyperref}
\usepackage[normalem]{ulem}

\begin{document}

\title{Non-linear current and dynamical quantum phase transitions \\in the flux-quenched Su-Schrieffer-Heeger model}
\author{Lorenzo Rossi}
\email{lorenzo.rossi@polito.it}
\affiliation{Dipartimento di Scienza Applicata e Tecnologia del Politecnico di Torino, I-10129 Torino, Italy}

\author{Fabrizio Dolcini}
\affiliation{Dipartimento di Scienza Applicata e Tecnologia del Politecnico di Torino, I-10129 Torino, Italy}

\begin{abstract} 
We  investigate the dynamical effects of a magnetic flux  quench in the Su-Schrieffer-Heeger model in a one-dimensional  ring geometry. We show that, even when the system is initially in the half-filled insulating state, the flux quench induces a time-dependent current that eventually reaches a finite stationary value. Such  persistent current, which  exists also  in the thermodynamic limit,  cannot be captured  by the linear response theory and is the hallmark of   nonlinear dynamical effects occurring in the presence of dimerization. Moreover, we show that, for a range of values of dimerization strength and initial flux, the system exhibits  dynamical quantum phase transitions, despite the quench is performed within the same topological class of the model.
\end{abstract}
\maketitle
\section{Introduction}
Many important features of a quantum mechanical system can be gained from the Linear Response Theory (LRT), where the out of equilibrium response of the system to a weak perturbation is encoded in a correlation function evaluated at its equilibrium state\cite{kubo_1957,mahan}. In particular, LRT is used to establish whether a fermionic system is a conductor or an insulator. Operatively this can be done through the following Gedankenexperiment: We first imagine to switch off all sources of extrinsic scattering phenomena, e.g. with a bath or with disorder. Then, we apply  a weak uniform electric pulse $E(t)=\mathcal{E} \delta(t)$ and observe the long time behavior of the current   in the thermodynamic limit. 
If a finite persistent  current eventually flows,  the system is  a conductor, otherwise it is an insulator.  
Explicitly,  the LRT current is expressed as $J(t)=(2\pi)^{-1}\mathcal{E}  \int d\omega\, \sigma(\omega)  e^{-i \omega t}$, and its  stationary value $J(+\infty)=D \mathcal{E}$ is determined by the  Drude weight $D$, i.e. the coefficient appearing in the low frequency singular term of  the conductivity\cite{resta_2018}  $\sigma(\omega)= \sigma_{\text{reg}}(\omega)+i D/(\omega+i0^+)$ and characterizing the possibility of a system to sustain   ballistic transport.
The evaluation of the Drude weight\cite{kohn_1964}   has allowed to identify interaction-induced insulating states in exactly solvable fermionic models, either by a direct investigation, like e.g. in the Hubbard model\cite{shastry_PRL_1990,kawakami_PRB_1991,zotos_1995} , or indirectly through spin models that can be mapped into fermionic ones through the Jordan-Wigner transformation\cite {zotos_1999,klumper_2005,zotos_2011,moore_2012,prosen_PRL_2014,prosen_2017}.
Moreover, the linear response of systems that are in a stationary out-of-equilibrium  state has been investigated \cite{rossini_2014}.

Remarkably, the high control and tunability of cold atom systems in optical lattices\cite{lewenstein_2007,bloch_2008}, together with the ability to realize artificial gauge fields\cite{dalibard_2011,goldman_2014}, intriguingly suggest that the above Gedankenexperiment could actually be realized  in a quantum quench protocol\cite{calabrese_2006,polkovnikov_review,eisert_2015,mitra_2018}. Consider an isolated fermionic system on a one dimensional (1D)  ring, initially prepared in the ground state of a given Hamiltonian $\hat{H}_i$. Then, suppose that the unitary dynamics is governed by a different final Hamiltonian $\hat{H}_f$, obtained from the previous one by a sudden change in a magnetic flux piercing the ring. Such sudden variation precisely generates the uniform electric pulse mentioned above.

These   experimental  advances  have also spurred the interest in   the  dynamics {\it beyond} LRT, i.e. when the stationary state properties of the system are no longer  sufficient  to describe its dynamical response. 
In particular, the dynamics resulting from a   flux quench has been   analyzed in the case of a single-band model of spinless fermions with a homogeneous nearest neighbour hopping and interaction\cite{misguich}. Although quantitative discrepancies from the LRT prediction have been numerically found in the gapless phase, the overall qualitative picture relating the existence of a persistent current to a non vanishing Drude weight seems quite robust.

In this paper we instead highlight qualitative differences from LRT predictions emerging after a flux quench in a model of spinless fermions hopping in a {\it dimerized} ring lattice. Specifically, we shall focus on the Su-Schrieffer-Heeger(SSH) model\cite{SSH_PRL1979,SSH_PRB1980}, recently realized in optical lattices\cite{bloch_2013,gadway_2016,gadway_2019}.
As is well known, such model is gapped   even without interactions and, at half filling,  describes a two-band (topological) insulator\cite{ungheresi_book,shen_book}. By quenching the initial flux to zero and by analyzing the resulting dynamics, we find two main results. First, while LRT predicts a vanishing Drude weight and a vanishing current\cite{resta_2018},   the flux quench does lead  to a persistent current flowing along the ring, which is thus a signature of non-linear effects. 
Second,  if the initial flux exceeds a critical value (dependent on the dimerization strength) dynamical quantum phase transitions (DQPTs)\cite{heyl_2013} occur. Notably, while a quench performed across the two different topological phases of the SSH model is known to give rise to DQPTs\cite{dora_2015}, the DQPTs we find occur even if the quench is performed within the same topological phase.

We emphasize that  the  effects predicted here are   intrinsically ascribed to the dimerization and arise even without interaction, 
in sharp contrast with the customary single-band tight-binding model with homogeneous hopping, where interaction is needed to observe any non trivial dynamical effect of the flux-quench\cite{misguich}. Here,   dimerization  provides an intrinsically  spinorial nature to the Hamiltonian and to its eigenstates,   implying
that the current operator is not a constant of motion even without interaction. Furthermore,  in the single band model the eigenstates of the Hamiltonian are uniquely determined by their (quasi)-momenta  and do not depend on the flux, while
in the dimerized case  the eigenstates exhibit  a non-trivial dependence on the flux. 
Finally, it is the spinorial nature, which is thus absent in the single-band tight-binding model, that leads to the DQPTs.

Our paper is organized as follows. In  Sec.\ref{sec-2} we present the model and describe the flux quench dynamics of a two-band model. In  Sec.\ref{Sec-Current} we derive the expression of the persistent current and show that, while in the limit of vanishing dimerization the LRT captures the metallic behavior, in the presence of dimerization the persistent current flows despite the LRT   predicts a vanishing Drude weight and an insulating behavior. In Sec.\ref{sec_DQPT} we then analyze the DQPTs induced by the dimerization. Finally, in  Sec.\ref{Sec-conclusions} we discuss our results and draw our conclusions.

\section{Model and state evolution}
\label{sec-2}
\subsection{The SSH model}
As mentioned in the Introduction, in this article we focus on a well known example of a band insulator, namely the  Su-Schrieffer-Heeger (SSH) model\cite{SSH_PRL1979,SSH_PRB1980}, in a 1D   ring  pierced by a magnetic flux. Here below we briefly recall a few aspects about this model that are needed to our analysis. The SSH Hamiltonian in real space is 
\begin{equation}\label{SSH-Ham-real}
    \hat{H}[\phi] = v \sum_{j=1}^M \left( e^{i \phi}\hat{c}_{j A}^\dagger \hat{c}_{j B} + r e^{i\phi}\hat{c}_{j B}^\dagger \hat{c}_{j+1 A} + \text{H. c.} \right)\\,
\end{equation}
where $M$ denotes the number of cells, containing two sites $A$ and $B$ each, $v$ is a real positive hopping amplitude, $r \ge 0$ is the dimerization parameter, and $\hat{c}_{j \alpha}^\dagger$ creates a spinless fermion in the site $\alpha=A,B$ of the $j$-th cell. Denoting by $\Phi$ the total magnetic flux threading the ring, we adopt the gauge where the phase related to its vector potential\cite{peierls,graf-vogl}, denoted by $\phi$ in Eq.(\ref{SSH-Ham-real}), is uniform along the ring links, so that  $2M\phi=2\pi (\Phi/\Phi_0)$, where $\Phi_0=h/e$ is the elementary flux quantum. We are interested in the thermodynamic limit $M\rightarrow +\infty$ with a finite flux per unit cell $\Phi /M$. 
 
In Eq.(\ref{SSH-Ham-real}) we assume periodic boundary conditions (PBCs), so that the $k$ wavevectors are quantized (also in the presence of flux) as $ka = 2\pi \, n/M$, where $n\in \left\{ {-\left\lfloor  M/2 \right \rfloor, \ldots , \left\lfloor (M-1)/2 \right\rfloor } \right\}$ and $a$ denotes the size of the unit cell. The 
 SSH Hamiltonian  is thus rewritten in momentum space as  
\begin{eqnarray}\label{SSH-Ham-k}
    \hat{H}[\phi] 
    &=& v\sum_{ka=-\pi}^{\pi} (\hat{c}^{\dagger}_{k A},\hat{c}^{\dagger}_{k B}) \,
     { \mathbf{d}(k,{\phi})  \cdot \boldsymbol\sigma} 
    \begin{pmatrix} \hat{c}^{}_{k A}\\ \hat{c}^{}_{k B}\end{pmatrix}
\end{eqnarray}
where 
\begin{eqnarray}
\lefteqn{    { \mathbf{d}(k,{\phi})}  \, \, =} & & \label{d-vec}\\
&=&   (  \cos{\phi} + r \cos(ka+{\phi})   ,    - \sin{\phi} + r \sin(ka+{\phi})   ,    0  )  \nonumber 
\end{eqnarray}
and $\boldsymbol\sigma =(\sigma_x,\sigma_y, \sigma_z)$ are Pauli matrices acting on the sublattice degree of freedom.
 The spectrum of single particle eigenvalues consists of two symmetric {energy bands $\varepsilon_\pm(k,{\phi})=\pm v \epsilon(k,{\phi})$, where
\begin{equation}\label{SSH-spectrum}
    \epsilon(k,{\phi})=   \sqrt{1+r^2+2r\cos(ka+2{\phi})} \quad . 
\end{equation}
The density matrices of the single particle eigenstates, in the $\{|kA\rangle,|kB\rangle\}$ basis, are given by
\begin{equation}\label{rho-SSH-eigenstates}
    \rho_\pm(k,{\phi})=\frac{1}{2}\left(\sigma_0\pm  { \mathbf{u}(k,{\phi})  \cdot  \boldsymbol\sigma } \right)
\end{equation}
where $\sigma_0$ is the $2\times2$ identity matrix, and $\mathbf{u}(k,{\phi})=\mathbf{d}(k,{\phi})/|\mathbf{d}(k,{\phi})|$ is a unit vector.

The SSH model is also known as a prototype model of a topological insulator\cite{ungheresi_book}, which exhibits two topologically distinct  phases for $r<1$ and $r>1$, with $r=1$ identifying the non-dimerized gapless case. 
Notably, in the presence of a magnetic flux  (${\phi} \neq 0$), the energy spectrum (\ref{SSH-spectrum}) depends on the wavevector $k$ and on the flux phase ${\phi}$ only though the combination $ka+2{\phi}$, whereas the Hamiltonian (\ref{SSH-Ham-k}) and its eigenstates (\ref{rho-SSH-eigenstates}) depend on {\it both} these quantities separately. 
This is due to the dimerization. Indeed, in the limit $r \rightarrow 1$ of vanishing dimerization,  in the Hamiltonian (\ref{SSH-Ham-k})  one has ${   \mathbf{d}(k,{\phi}) \cdot  \boldsymbol\sigma}=2\cos(k a/2 + {\phi}) (\sigma_x \cos(ka/2)+\sigma_y \sin(ka/2))$,  and the dependence on the flux phase reduces to a mere multiplicative factor. In this case  the single particle eigenstates  become independent of ${\phi}$.

\subsection{State evolution upon a flux quench}
Let us suppose that the system is initially prepared in the  insulating  ground state of the half filled  SSH model with an initial flux phase value ${\phi}_i$,  corresponding to a completely filled lower band {$\varepsilon_-(k,\phi_i)$}.   The $k$-th component of the single particle density matrix at $t=0$ can thus be written in the $\{|kA\rangle,|kB\rangle\}$ basis as $\rho_i(k)=\big[\sigma_0-\mathbf{u}_i(k) \cdot \boldsymbol\sigma \big]/2$, where $\mathbf{u}_i(k)= \mathbf{u}(k,{\phi}_i)$.  
Then,   the magnetic flux is suddenly switched off and  the initial state evolves according to the final Hamiltonian $\hat{H}_f$ characterized in Eq.(\ref{SSH-Ham-k}) by $\mathbf{d}_f(k)= \mathbf{d}(k,{\phi}=0)$, which in turn identifies the unit vector $\mathbf{u}_f(k)= \mathbf{u}(k,{\phi}=0)$. 
 
Since the $k$ modes do not couple in the quench process, the Liouville-Von Neumann equation can be easily integrated and the $k$-th component of the one-body density matrix
is  uniquely identified, in the $\{|kA\rangle,|kB\rangle\}$ basis, by the time evolving Bloch vector $\mathbf{u}(k,t)$ through
\begin{equation}\label{rho-evolution}
    \rho(k,t)=\frac{1}{2}\big[\sigma_0-\mathbf{u}(k,t) \cdot \boldsymbol\sigma\big] \quad.
\end{equation}
Specifically, the Bloch vector precesses around the final direction $\mathbf{u}_f(k)$ and can be expressed as   the sum of three orthogonal contributions\cite{yang_2018}
\begin{eqnarray}
\lefteqn{\mathbf{u}(k,t)= } & & \label{Bloch-vec-evolution} \\
&=& \mathbf{d}_{\parallel}(k)+ \mathbf{d}_{\bot}(k) \cos \big[\frac{2  {\epsilon(k,0)}   vt}{\hbar}\big]+\mathbf{d}_{\times}(k) \sin \big[\frac{2  {\epsilon(k,0)}    vt}{\hbar}\big] \nonumber
\end{eqnarray}
whose explicit expressions can be deduced from the  general state evolution in a two-band model (see Appendix \ref{App-two-band-state-evolution}) and read
\begin{eqnarray}
 \mathbf{d}_{\parallel}(k ) \, \, &=& 
          d_{\parallel}(k ,{\phi}_i)  \mathbf{u}_{f}(k) \label{d-parallel} \\ \nonumber
          \\
 \mathbf{d}_{\bot}(k) &=& d_{\bot}(k ,{\phi}_i)   \mathbf{R}_z [ \mathbf{u}_{f}(k) ] \label{d-perp}  \\ \nonumber
          \\
 \mathbf{d}_{\times}(k)&=& d_{\bot}(k ,{\phi}_i) (-{\mathbf{e}}_z)  \label{d-z} \quad.
\end{eqnarray}
Here 
\begin{equation}
    \mathbf{R}_z=
    \begin{pmatrix} 
    0 & 1 & 0 \\
    -1 & 0 & 0 \\
    0 &  0 & 1 
    \end{pmatrix}
\end{equation}
is a matrix describing a rotation by $-\pi/2$ around the $z$-axis identified by the unit vector $\mathbf{e}_z$ and orthogonal to the $\mathbf{d}_{i}$-$ \mathbf{d}_{f}$ plane, while 
\begin{eqnarray}
d_{\parallel}(k ,{\phi}_i)  &=& \frac{(1+r^2)\cos{\phi}_i + 2 r \cos(ka+{\phi}_i)}
    {{\epsilon(k,\phi_i)}{\epsilon(k,0)}} \label{dpara} \\
d_{\bot}(k ,{\phi}_i)  &=& \frac{(1-r^2)\sin{\phi}_i }
    {{\epsilon(k,\phi_i)}{\epsilon(k,0)}}   \label{dperp} \quad.
\end{eqnarray} 
As a last remark we notice that, in the  limit $r\rightarrow 1$ of vanishing dimerization,   the dynamics in Eq.(\ref{Bloch-vec-evolution}) becomes trivial, since  
$d_{\bot}(k,\phi_i)=0$ and $ d_{\parallel}(k ,\phi)=\text{sign}[\cos(ka/2+\phi_i)\cos(ka/2)]$.
Indeed  without dimerization the initial state is an eigenstate of $\hat{H}_f$ and its density matrix does not evolve with time.\\

\section{Current}\label{Sec-Current} 
Let us now investigate the dynamical behavior of the particle current generated by the quench. We first note that, because the system is bipartite, there actually exist two types of currents, namely inter-cell and to intra-cell current operators. Their explicit expression straightforwardly stems from the continuity equation related to the post-quench Hamiltonian~$\hat{H}_f$ (see Appendix \ref{App-Current}) and reads  
\begin{eqnarray}
    \hat{J}_j^{inter}&=&\frac{rv}{\hbar} \left[ i \hat{c}_{jB}^\dagger \hat{c}_{j+1A}- i \hat{c}_{j+1A}^\dagger \hat{c}_{jB}\right]  \label{Jjinter-def} \\
    \hat{J}_j^{intra}&=&\frac{v}{\hbar} \left[ i  \hat{c}_{jA}^\dagger \hat{c}_{jB}- i \hat{c}_{jB}^\dagger \hat{c}_{jA}\right] \label{Jjintra-def} \quad . 
\end{eqnarray}
Note that, since $\hat{H}_f$ has a vanishing flux, these operators do not depend on the flux explicitly. Due to the translational invariance of both the initial state and the final Hamiltonian, the expectation values of Eqs.(\ref{Jjinter-def})-(\ref{Jjintra-def})  are actually independent on the specific cell label $j$. It is thus worth introducing   the  space-averaged operators $\hat{J}^{l}\equiv  M^{-1}  \sum_{j=1}^M \hat{J}_j^{l}$ (with $l=inter/intra$), obtaining
\begin{equation}
 \hat{J}^{l}   =     \frac{1}{M} \sum_{ka=-\pi}^\pi  (\hat{c}^{\dagger}_{k A},\hat{c}^{\dagger}_{k B}) \,
    \mathcal{J}^l_k
    \begin{pmatrix} \hat{c}^{}_{k A}\\ \hat{c}^{}_{k B}\end{pmatrix}  
\end{equation}
where
\begin{eqnarray}
 \mathcal{J}^{inter}_k   & = &  \frac{rv}{\hbar} \left( -\sin(k a) \sigma_x +\cos(ka) \sigma_y\right) \label{Jinter-1body} \\ 
\mathcal{J}^{intra}_k   & = & - \frac{v}{\hbar}   \sigma_y \quad.\label{Jintra-1body} 
\end{eqnarray}
Their expectation values $J^l(t) \equiv \langle \hat{J}^{l}\rangle(t)=M^{-1}\sum_{k}{\rm tr}[\mathcal{J}^{l}_k \rho(k,t)]$ {for $t>0$} can be written as
\begin{equation}\label{Jl(t)}
J^l(t)=J_{dc} + J^l_{ac}(t) \hspace{1cm}l=inter/intra
\end{equation}
where the first term $J_{dc}$ describes a steady state contribution and is thus the same for inter/intra contributions, while the second term describes the time-dependent  fluctuations around it and {is different in  the two contributions}. Explicitly, the {\it ac}-{terms}  read 
\begin{eqnarray}
   J^{inter}_{ac}(t) &=&    \frac{rv}{\hbar} \frac{1}{M} \sum_{ka=-\pi}^\pi  
        d_\bot(k,{\phi}_i) 
        \frac{
        r+ \cos(ka)}
        {{\epsilon(k,0)}}
        \, \, \times \nonumber \\
        & & 
        \hspace{2.cm} \times \cos\big[\frac{2{\epsilon(k,0)} vt}{\hbar}\big] 
\end{eqnarray} 
and
\begin{eqnarray}        
J^{intra}_{ac}(t)  & = & -  \frac{v}{\hbar} \frac{1}{M} \sum_{ka=-\pi}^\pi  
       d_\bot(k,{\phi}_i) 
        \frac{
        1+ r\cos(ka)}
        {{\epsilon(k,0)}}
        \, \, \times \nonumber \\
        & & 
        \hspace{2.cm} \times \cos\big[\frac{2{\epsilon(k,0)} vt}{\hbar}\big] 
\end{eqnarray}
whereas the {\it dc}-current is 
\begin{equation} \label{J-dc}
   J_{dc}  =    \frac{rv}{\hbar} \frac{1}{M} \sum_{ka=-\pi}^\pi 
       d_\parallel(k,{\phi}_i) 
        \, \,
        \frac{
         \sin(ka)}
        {{\epsilon(k,0)}}
\end{equation}     
with $d_\parallel (ka,{\phi}_i)  $ and $d_\bot (ka,{\phi}_i) $   given 
by Eqs.(\ref{dpara})-(\ref{dperp}). 
Figure \ref{J_vs_t} displays the time evolution of  $J^{intra}(t)$ and $J^{inter}(t)$ in the thermodynamic limit $M^{-1}\sum_k \rightarrow  (2\pi)^{-1}   \int d(ka)$. {As one can see, the two currents are in general different and exhibit  long living fluctuations, described by the {\it ac}-terms in Eq.(\ref{Jl(t)}). However, these fluctuations eventually vanish and both currents converge to the same steady state contribution $J_{dc}$, highlighted by the green line.}

\begin{figure}[]
\centering
\includegraphics[width=\columnwidth,clip]{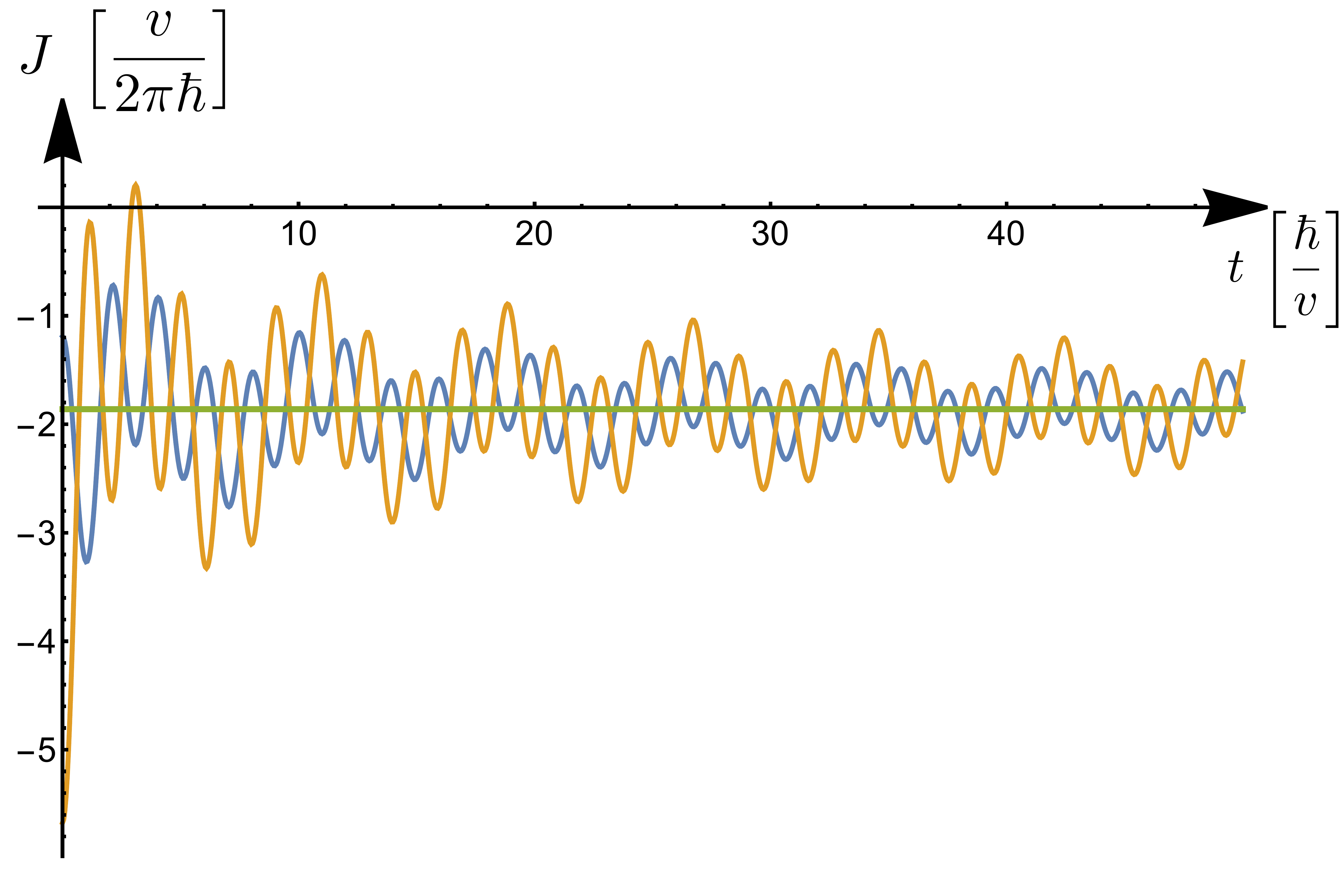}
\caption{ The inter-cell current $J^{inter}$ (blue curve) and the intra-cell current $J^{intra}$ (yellow curve)  resulting from  a sudden flux quench    in the SSH model  are plotted as a function of time. At  long-time, they both tend to the same stationary contribution $J_{dc}$ (green curve). The time evolution is computed in the thermodynamic limit for $r=0.6$ and $\phi_i= \pi/2$.    }
\label{J_vs_t}
\end{figure}

A few comments are in order about such persistent current $J_{dc}$.
First, $J_{dc}$ is essentially different from the   current  flowing at equilibrium in a mesoscopic ring threaded by a flux,  since it is  non-vanishing also in the thermodynamic limit, where it acquires the form
\begin{eqnarray}
J_{dc} &=&   { \frac{v}{ 2\pi\hbar }} \int_{-\pi}^\pi   d(ka) 
   \,
    \frac{
        r \sin(ka)}
        { {\epsilon(k,0)}}  \times \nonumber \\
        & & \times \frac{(1+r^2)\cos{\phi}_i + 2 r \cos(ka+{\phi}_i)}
    {{\epsilon(k,\phi_i)}{\epsilon(k,0)}} \label{Jdc-TL}
\end{eqnarray}
Second, $J_{dc}$ cannot be captured by the LRT, which would predict a vanishing persistent current due to a vanishing Drude weight (see Appendix \ref{App-Drude}).  This can also be seen by inspecting  Eq.(\ref{Jdc-TL}) in the limit of weak initial flux ${\phi}_i\ll1$,  which corresponds to the limit of weak applied electric pulse. Indeed one obtains 
\begin{eqnarray}\label{Jdc-insulating}
      J_{dc}  
     \approx 
    - { \frac{v \,r^2}{\pi\hbar }}  (1-r^2)^2 \left[ \int_{-\pi}^\pi   d(ka) 
    \frac{\sin^2(ka)}{{\epsilon^7}(k,0)} \right] {\phi}_i^3 \,\,
\end{eqnarray}
which highlights the non-linear (cubic) response of the insulating SSH ring. 

It is  now worth comparing the above results with the one of the non-dimerized  limit $r \rightarrow 1$, where one obtains {for the post-quench currents ($t>0$)}
\begin{eqnarray}
J^{inter}(t)=J^{intra}(t)&=&
    -{\frac{2v}{ \pi \hbar} }   \sin{\phi}_i  \quad .\label{Jdc-metallic}
\end{eqnarray}
Differently from the result obtained for the dimerized case (see Fig.\ref{J_vs_t}), the  current (\ref{Jdc-metallic}) is time-independent after the quench\cite{footnote}  and, for a weak field $ \phi_i \ll 1$, it exhibits a linear dependence on $\phi_i$. One thus recovers the well known finite Drude weight\cite{footnote_Drude} $ D = - (e^2/\hbar) \mathtt{v}_F/\pi $, where $\mathtt{v}_F$ is the Fermi velocity, of a non interacting half filled metallic band, as predicted by LRT\cite{giamarchi_book}.
\\
The role of dimerization is emphasized in Fig.\ref{Jdc_vs_phi}, where the persistent current  (\ref{Jdc-TL}) is depicted as a function of the initial flux, for various values of dimerization~$r$. While at small flux values $\phi_i \ll 1$ the current   $J_{dc}$ of the dimerized case $r \neq 1$ is suppressed as compared to the metallic case $r=1$ (green curve), for finite flux values the two cases exhibit comparable currents. 
\begin{figure}[b]
\centering
\includegraphics[width=\columnwidth,clip]{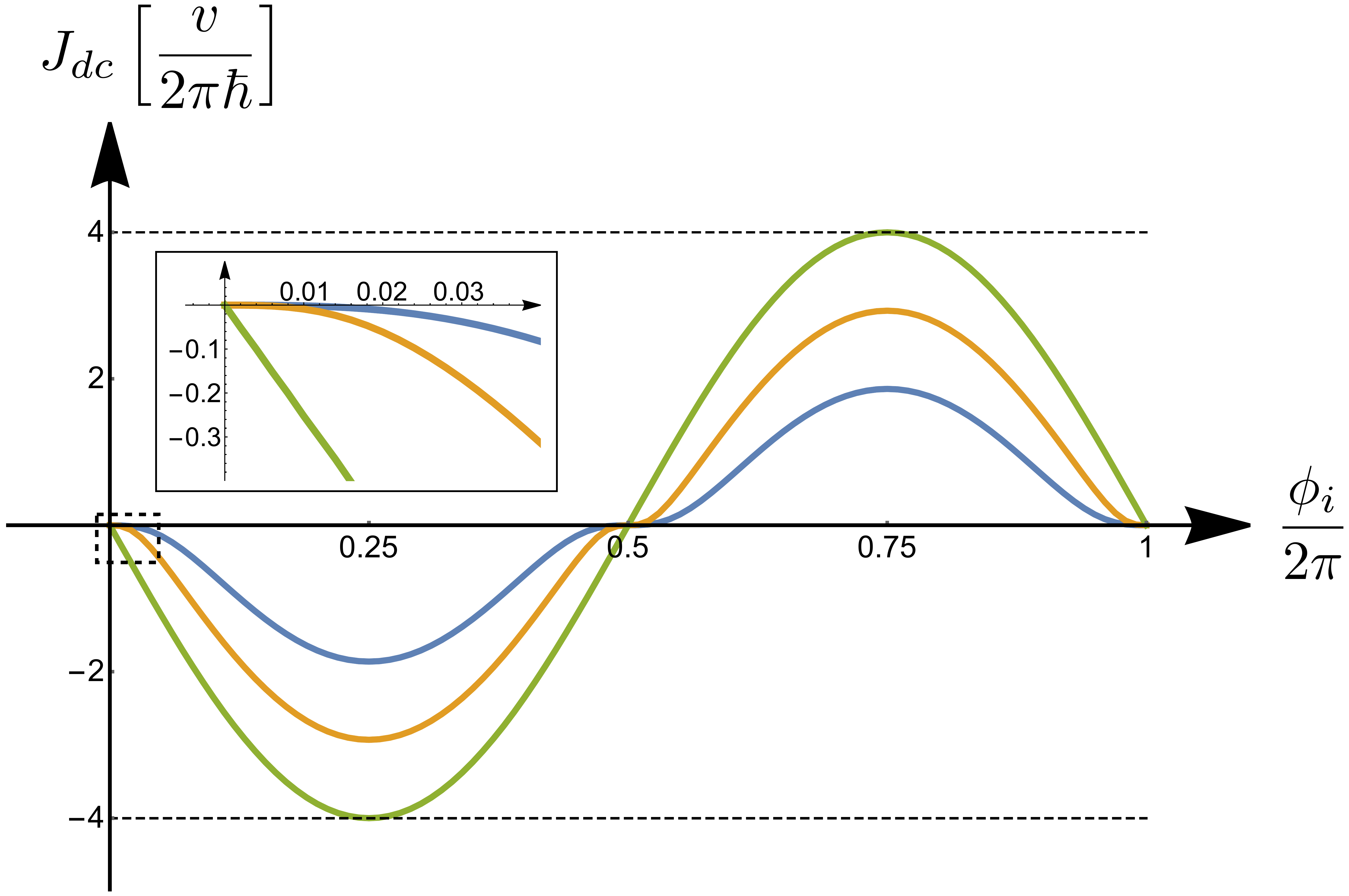}
\caption{The {persistent current $J_{dc}$} induced in  the SSH model  by  quenching the flux to zero  is plotted as a function of the initial flux $\phi_i$.  The blue, yellow and green  curves are obtained for different dimerization strengths, namely $r=(0.6,0.8,1)$ respectively. For each value of $r\ne 1$ the current does not exhibit a linear term in $\phi_i$ for $\phi_i \ll 1$.  The inset magnifies the behaviour at small fluxes to highlight the difference between linear and non linear response. }
\label{Jdc_vs_phi}
\end{figure}

The  origin of the persistent current term $J_{dc}$  can be understood in terms of  the out of equilibrium  occupancies $n_{f,\pm}$ of the post-quench  bands $\varepsilon_\pm(k,0)$ induced by the flux quench. These can be computed, for each $k$, by projecting the initial state on the post-quench eigenmodes, obtaining time-independent expressions
\begin{eqnarray}
\lefteqn{ n_{f,\pm}(k,{\phi}_i)  = \text{tr} \left\{ \rho_i(k)   \left( \sigma_0 \pm  \mathbf{u}_f(k) \cdot \boldsymbol\sigma  \right)/2  \right\} } & &  \nonumber \\
     &=& \frac{1}{2} (1 \mp \mathbf{u}_i(k)\cdot \mathbf{u}_f(k) )= \nonumber\\
     &=& \frac{1}{2}  \mp
     \frac{(1+r^2)\cos{\phi}_i + 2 r \cos(ka+{\phi}_i)}
     {2 {\epsilon(k,{\phi}_i) 
          \epsilon(k,0)}}    \label{post-quench-occupancies}
    \end{eqnarray}
which are plotted as a function of $ka/\pi$ in Fig.\ref{occupancy}.
By comparing Eq.(\ref{post-quench-occupancies}) with Eq.(\ref{Jdc-TL}),  the persistent current can be rewritten as 
\begin{eqnarray}
 J_{dc} 
 =\frac{1}{2\pi a} \int_{-\pi}^\pi d(ka) \,    
     \Delta n_f(k,{\phi}_i) \,
\frac{1}{\hbar} \partial_k \varepsilon_-(ka)  \label{Jdc-expr2}
\end{eqnarray}
where 
 $\frac{1}{\hbar} \, \partial_k \varepsilon_\pm(k)=\pm \frac{v a}{\hbar}  \frac{r \sin(ka)}{ \epsilon(ka)}$ are the post quench group velocities, and
\begin{eqnarray}
 \Delta {n}_f(k,{\phi}_i) & =&   {n}_{f,-}(k,{\phi}_i) -{n}_{f,+}(k,{\phi}_i) = \nonumber \\
 &=& \mathbf{u}_i(k) \cdot \mathbf{u}_f(k)  \label{Delta-nf}
\end{eqnarray}
denotes the occupancy difference.  
{Since in Eq.(\ref{Jdc-expr2}) the group velocities are odd functions in $k$,  the origin of the non-vanishing persistent current $J_{dc}$ boils down to the lack of even parity in $k$ of the  post quench occupancy distributions (\ref{post-quench-occupancies}) and of their difference $ \Delta {n}_f$. Such lack of symmetry, clearly seen in  Fig.\ref{occupancy}, arises from the fact that the flux quench impacts on the {\it phase} of tunneling amplitudes, whereas quenches in the magnitude of the tunneling amplitudes lead to   out of equilibrium occupancy distributions that preserve their even parity in $k$ and cannot induce  a net current~\cite{porta_2018}.}
\begin{figure}[]
\centering
\includegraphics[width=\columnwidth,clip]{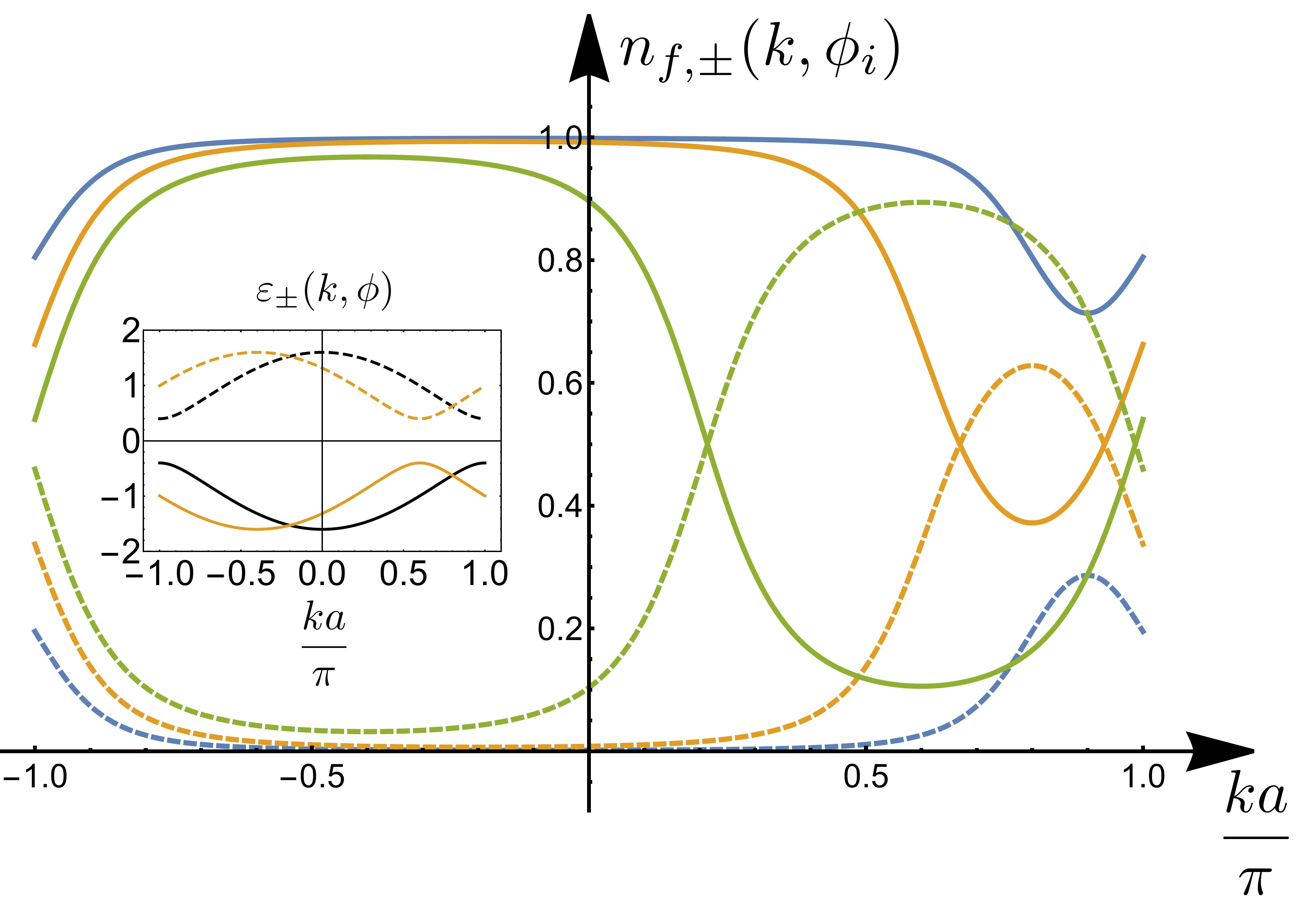}
\caption{Occupancies of the post-quench bands for different values of the initial flux $\phi_i$ and a fixed dimerization strength $r=0.6$. Dashed lines correspond to the upper band, while solid lines correspond to the lower one. The blue, yellow and green colors correspond to $\phi_i=(0.1,0.2,0.4)\pi$, respectively. The distributions are not symmetric  in $k\leftrightarrow-k$ for any value of the initial flux. As $\phi_i$ is increased, the  upper band becomes more occupied and the lower band gets more depleted.  Inset: bands {$\varepsilon_\pm(k,\phi)$} of a SSH model pierced by a magnetic flux: Solid and dashed lines describe the lower and the upper bands, respectively. The bands are depicted for  {$v=1$}, a fixed dimerization strength $r=0.6$ and for two different values of the flux, namely $\phi=0.2\pi$ (yellow lines) and $\phi=0$ (black lines). }
\label{occupancy}
\end{figure}

We conclude this section by two comments. First, when moving away from half filling, the system becomes metallic even in the presence of dimerization. In this case one can show that the system develops a finite Drude weight and that  the  linear response theory well captures the quench-induced current for small initial fluxes. Nonetheless, there exist some qualitative differences with respect to the non-dimerized  metallic case. Indeed, because of dimerization, the current also has a finite {\it ac}-contribution and, for small filling, it does not increase monotonically in $\phi_i \in [0, \pi/2]$, developing a local minimum for $\phi_i=\pi/2$ instead of a maximum. 
The second comment is concerned with the flux switching protocol. Here, in analogy to what was done in Ref.[\onlinecite{misguich}], we have considered the switching off of the initial flux, so that the latter only appears in the initial state. In the reversed protocol, where the flux is switched on, one  obtains a current with opposite sign, as expected, provided that one consistently includes the flux phases related to the vector potential both in the post-quench Hamiltonian and in the current operators (\ref{Jjinter-def})-(\ref{Jjintra-def}).

\section{Dynamical Quantum Phase Transitions}\label{sec_DQPT}
Let us now analyze the properties of the Loschmidt amplitude $\mathcal{G}(t)=\langle \psi_0 | e^{- i \hat{H}_f t / \hbar} | \psi_0 \rangle $, where $|\psi_0\rangle$ is the many-body initial state, while $\hat{H}_f$ is the final Hamiltonian that governs the time evolution of the system after the quench.  With applications in studies on quantum chaos and dephasing \cite{jalabert_2001,jarzynski_2002,zanardi_2006}, the Loschmidt amplitude has a tight relation to the statistics of the work  performed through the quench \cite{silva_2008,heyl_2013,deluca_2014}. Equivalently, it can also be regarded as the generating function of the energy probability distribution encoded in the post quench diagonal ensemble, since $\mathcal{G}(t)=\int dE P(E) e^{-i Et/\hbar}$ and the post-quench diagonal ensemble is described by $P(E)=\sum_{n} | \langle n  | \psi_0 \rangle |^2 \delta(E-E_n)$, where $E_n$ and $| n \rangle$ are the many-body eigenvalues and eigenstates of the final Hamiltonian respectively. 
Moreover, it has been suggested\cite{heyl_2013}   that the Loschmidt amplitude can be interpreted as a dynamical partition function whose zeros, in analogy with the equilibrium case, are identified with DQPTs. The initial belief of a connection between DQPTs and quenches across different equilibrium phase transitions has been proved to be not rigorous \cite{dora_2014,andraschko_2014,jafari_2017,jafari_2019,budich_2020,jafari_2021}, 
and the impact of DQPTs on local observables has been found only in specific cases\cite{heyl_2013,fagotti_2013,halimeh_2017,halimeh_2017bis, halimeh_2018,silva_2018}.
Nevertheless, the existence of zeros of $\mathcal{G}(t)$ can be interpreted as a clear signature of  quench-induced population inversion\cite{heyl_2013,defenu_2019}. 

\begin{figure}[]
\centering
\includegraphics[width=\columnwidth,clip]{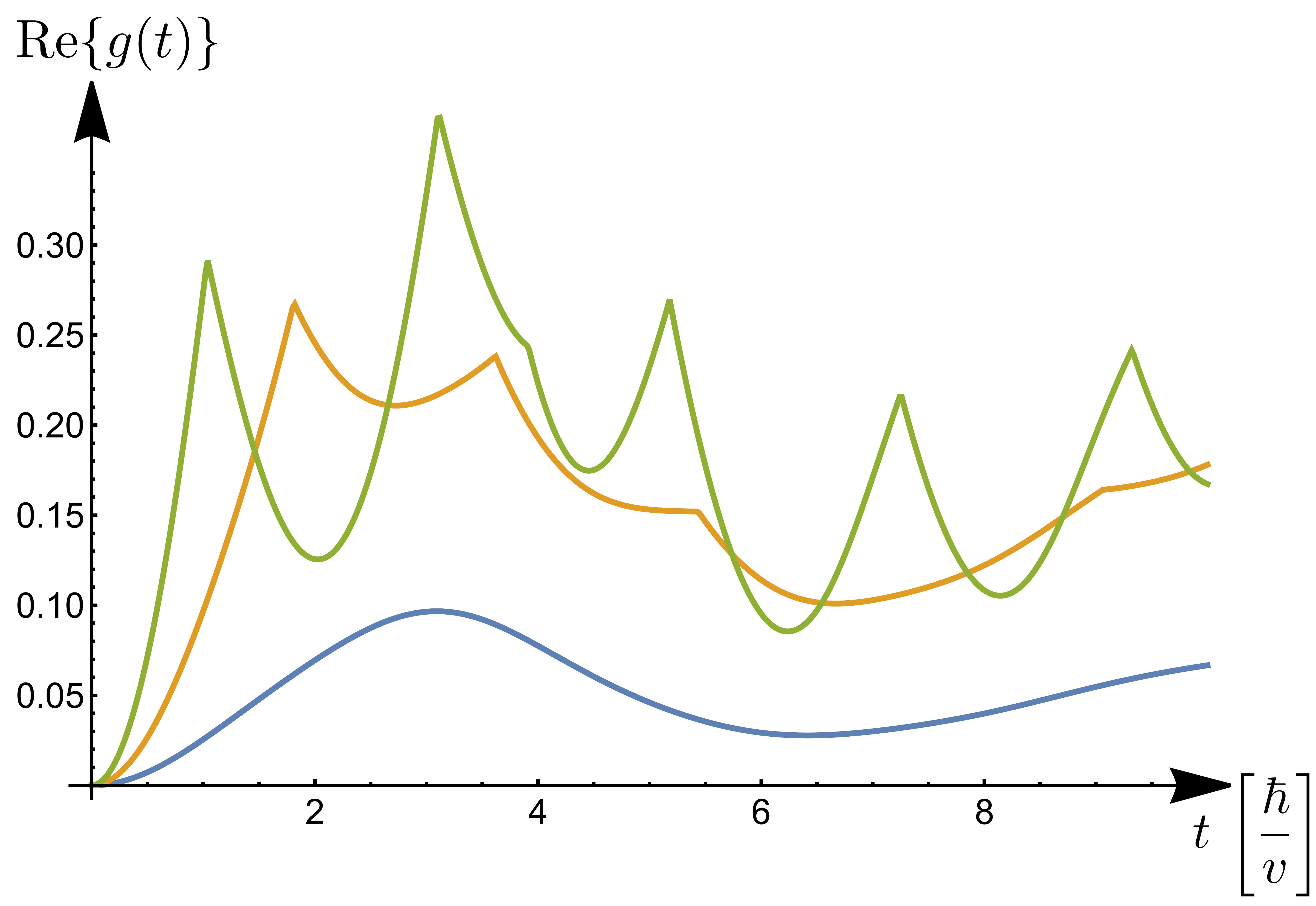}
\caption{Time evolution of the real part of the  dynamical free energy density $g(t)$, for different values of the initial flux $\phi_i$ at a fixed dimerization strength $r=0.6$. The blue curve   corresponds to an initial flux $\phi_i= 0.1 \pi$ lying outside the range  identified by Eq.(\ref{constraint})   and is smooth. The   yellow and green curves  correspond to flux values  that fulfill Eq.(\ref{constraint})  ($\phi_i=(0.2,0.4)\pi$, respectively) and exhibit DQPT singularities. }
\label{DQPT}
\end{figure}

For the present flux quench the Loschmidt amplitude explicitly   reads\cite{dora_2015} 
\begin{eqnarray}
\mathcal{G}(t)&=&\prod_{-\pi \le ka \le \pi}  \left[ \cos \left(\frac{{\epsilon(k,0)} v t}{\hbar}\right) \right.+ \nonumber \\  
& & \qquad \quad  \, \,  \left. +i \, [ \mathbf{u}_i(k) \cdot \mathbf{u}_f(k) ] \sin\left(\frac{{\epsilon(k,0)} v t}{\hbar}\right)  \right]
\end{eqnarray}
whence the dynamical free energy density $g(t)=-M^{-1} \log[\mathcal{G}(t)]$ in the thermodynamic limit    is straightforwardly given by 
\begin{eqnarray}\label{dyanmical free energy}
g(t) &= & -\frac{1}{2\pi} \int_{-\pi}^\pi d(ka) \log \left[ \cos\left(\frac{{\epsilon(k,0)} v t}{\hbar}\right) + \right. \nonumber \\
&& \qquad \quad \, \,  \left. +i  [\mathbf{u}_i(k) \cdot \mathbf{u}_f(k)] \sin\left(\frac{{\epsilon(k,0)} v t}{\hbar}\right) \right] \,.\,\,\label{g(t)}
\end{eqnarray}
The argument of the logarithm in Eq.(\ref{g(t)}) may vanish at some critical times if and only if
$
\mathbf{u}_i(k) \cdot \mathbf{u}_f(k)=0 
$.
Using Eqs.(\ref{Delta-nf}) and (\ref{post-quench-occupancies}) {in the regime $r\ne1$}, this condition can be satisfied by some $ka \in [-\pi,\pi]$ if and only if
\begin{equation}\label{constraint}
  | \cos\phi_i | \le \frac{2r}{1+r^2}\quad. 
\end{equation}
In conclusion, for each value of the dimerization strength $r\ne1$, there exists a   range of  initial flux values, Eq.(\ref{constraint}),  such that singularities in the dynamical free energy density appear, as shown in Fig.\ref{DQPT}. Recalling Eq.(\ref{Delta-nf}), we observe that DQPTs appear if and only if the post-quench band occupancies cross at some $k$, i.e. if there exists  a subregion of the Brillouin zone, where the post-quench upper band is more populated then the lower one (band population inversion). This is the case for the yellow and green curves in Figs.\ref{occupancy} and  \ref{DQPT}.  
Notably, while a quench across the critical point $r=1$ is  sufficient to induce a DQPT\cite{dora_2015}, it is not a necessary condition  and   \textit{accidental} DQPTs can also appear\cite{dora_2014,budich_2020}. This is the case here, where the DQPTs show up even if  the quench is performed within the same topological phase.

Before concluding this section, a remark is in order about the specific case $r=1$, which deserves some care. At first, by looking at the limit $r \rightarrow 1$ of Eq.(\ref{constraint}), one could naively expect that DQPTs exist for any value of the initial flux. However, this is not the case since the scalar product $\mathbf{u}_i(k) \cdot \mathbf{u}_f(k)$ reduces to a pure sign and the argument of the logarithm in Eq.(\ref{dyanmical free energy}) can never vanish. Indeed for $r=1$  the initial state is an (excited) eigenstate of the final Hamiltonian, its dynamics is trivial and $\mathcal{G}(t)$ reduces to a pure oscillating phase\cite{deluca_2014}. Hence the Loschmidt amplitude can never vanish and the dynamical free energy density is analytic for $t>0$.
Moreover, for $r=1$ a description in terms of a two band structure is redundant and a proper band population inversion can not be defined without ambiguities.

\section{Discussion and Conclusions}
\label{Sec-conclusions}
Our results  have been obtained in the case of  a sudden flux quench. Here we would like to briefly discuss the effects of a finite switch-off time $\tau_{sw}$.  By implementing a time-dependent flux phase $\phi(t)=\phi_i(1-\text{Erf}(\sqrt{8} t/\tau_{sw}))$    and by numerically integrating the Liouville von-Neumann equation for the density matrix, one can show that the persistent current $J_{dc}$  depends on the ratio $\tau_{sw}/\tau_g$, where $\tau_{g}=\hbar/(2v|1-r|)$ is the timescale associated to the energy gap of the SSH model. In particular, while for $\tau_{sw} \ll \tau_g$ the persistent current $J_{dc}$ is robust, when $\tau_{sw}\simeq \tau_g$ it  reduces with respect to the sudden quench value (e.g. to roughly $1/5$ for the parameters of Fig.\ref{J_vs_t}) and it vanishes in the limit $\tau_{sw} \gg \tau_g$ of an adiabatic switch-off.
In such limit, a vanishing stationary current    is consistent with the recent generalization of LRT to higher order response, which predicts that in a band insulator the  response to an adiabatic electric field vanishes to all orders in the field strength\cite{oshikawa1,oshikawa2,resta_arxiv}. 

It is worth pointing out the essential difference between the quench induced dynamics in an  insulating and in  a metallic state. For a metallic state, where the response to a weak electric pulse is linear, the  persistent current  that eventually flows  is independent of the quench protocol and is thus fully encoded in    the Drude weight. In striking contrast, when a weak field is applied to an   insulating state (like the half-filled SSH model),   the response is {\it non-linear} and does depend on the quench protocol. Thus, while the vanishing higher order generalized Drude weights\cite{oshikawa1,oshikawa2} only capture the behavior of the system in the adiabatic switching limit,   for a sufficiently fast switching a persistent current does flow even in an insulator.

In conclusion, in this paper we have analyzed the response of a half filled SSH ring to a sudden flux quench, or equivalently, to a sudden pulse of electric field. We have shown that the intrinsically spinorial nature of the problem, due to the dimerization  of the hopping amplitudes, induces a non trivial current dynamics even without interactions. In particular, a time-dependent current flows along the ring and  eventually reaches a stationary value, despite the insulating nature of the initial state (see Fig.\ref{J_vs_t}). Such persistent current $J_{dc}$,  which depends cubically on a weak initial flux $\phi_i$ in the presence of dimerization [see Eq.(\ref{Jdc-insulating}) and Fig.\ref{Jdc_vs_phi}], is a clear hallmark of a non-linear dynamics and it is ascribed to the peculiar non-equilibrium occupancy induced by the quench   (see Fig.\ref{occupancy}). 
For suitable dimerization and flux values,  a post quench   population inversion occurs, which in turn implies the occurrence of  DQPTs (see Fig.\ref{DQPT}). Notably, the DQPTs are present even without closing the gap, i.e. when the quench is performed within the same topological phase.

\acknowledgments
Fruitful discussions with Giuseppe Santoro, Rosario Fazio, Mario Collura and Alessandro Silva are greatly acknowledged.

\section{Appendix}

\subsection{State evolution in a quenched two-band system}
\label{App-two-band-state-evolution}
In this appendix we   recall the general state evolution after a sudden quench in a two-band model\cite{yang_2018}.  Let us  suppose that the initial state is the half filled ground state of a two-band Hamiltonian, whose one-body form can be written in momentum space  as 
\begin{equation}
    H_i(k)=v \left[ d_i^0(k) \sigma_0 + \mathbf{d}_i(k) \cdot \boldsymbol\sigma \right]  
\end{equation}
where $v$ denotes the reference energy scale.  The $k$-th component of the initial state can thus be written as $\rho(k,0)=\big[\sigma_0-\mathbf{u}_i(k) \cdot \boldsymbol\sigma\big]/2$
where $\mathbf{u}_i(k)=\mathbf{d}_i(k)/|\mathbf{d}_i(k)|$. The state evolves according to the post-quench Hamiltonian 
\begin{equation}
    H_f(k)=v\left[d_f^0(k) \sigma_0 + \mathbf{d}_f(k) \cdot \boldsymbol\sigma \right]  
\end{equation}
 and, by solving the Liouville-Von Neumann equation for the one-body density matrix, one can write the $k$-th component of the time evolved state as  $\rho(k,t)= \big[\sigma_0-\mathbf{u}(k,t) \cdot \boldsymbol\sigma\big]/2$, 
where the time dependent unit vector can be written as the sum of three orthogonal contributions:
\begin{eqnarray}
\mathbf{u}(k,t) &=& \mathbf{d}_{\parallel}(k)+ \mathbf{d}_{\bot}(k) \cos\big[2 |\mathbf{d}_f(k)| vt/\hbar\big]+ \nonumber \\
&& \hspace{27pt} +\mathbf{d}_{\times}(k) \sin\big[2 |\mathbf{d}_f(k)| vt/\hbar\big]  
\end{eqnarray}
where
\begin{eqnarray}
\mathbf{d}_{\parallel}(k) &=& \Big[ \mathbf{u}_{i} (k) \cdot \mathbf{u}_{f}(k) \Big] \mathbf{u}_{f}(k) \nonumber \\
\mathbf{d}_{\bot}(k)&=&\Big[ \mathbf{u}_{i}(k) - \mathbf{d}_{\parallel}(k) \Big] \nonumber  \\
\mathbf{d}_{\times}(k)&=&-\Big[ \mathbf{u}_{i} (k) \times \mathbf{u}_{f}(k) \Big]  \nonumber
\end{eqnarray}
Then, by inserting the explicit expression for $\mathbf{d}_{i}(k)$ and $\mathbf{d}_{f}(k)$ corresponding to a flux quench in the SSH model, see Eq. (\ref{d-vec}) of the Main Text, the expressions in Eqs. (\ref{d-parallel}), (\ref{d-perp}) and (\ref{d-z}) are recovered.

\subsection{Current} \label{App-Current}
In this appendix we  briefly outline how to derive the current operators discussed in Sec.\ref{Sec-Current}. Given the site density operators $\hat{n}_{j,\alpha}=\hat{c}_{j\alpha}^\dagger\hat{c}_{j\alpha}$, with $\alpha=A,B$, and the SSH Hamiltonian with $\phi=0$ (see Eq.(\ref{SSH-Ham-real}) of the Main Text), it is straightforward to derive the following Heisenberg equations of motion:
\begin{eqnarray}
        \partial_t \hat{n}_{jA}&=&  {\hat{J}_{j-1}^{inter}} - {\hat{J}_{j}^{intra}}  \nonumber \\
        \partial_t \hat{n}_{jB}&=&  {\hat{J}_{j}^{intra}} - {\hat{J}_{j}^{inter}} \nonumber
\end{eqnarray}
where, by definition:
\begin{equation}
        \hat{J}_{j}^{inter}=\frac{rv}{\hbar} \left[ i \hat{c}_{jB}^\dagger \hat{c}_{j+1A} - i \hat{c}_{j+1A}^\dagger \hat{c}_{jB} \right] \nonumber
\end{equation}
is the inter-cell current reported in Eq.(\ref{Jjinter-def}), while:
\begin{equation}
        \hat{J}_{j}^{intra}=\frac{v}{\hbar} \left[ i \hat{c}_{jA}^\dagger \hat{c}_{jB}-i \hat{c}_{jB}^\dagger \hat{c}_{jA} \right] \nonumber
\end{equation}
is the intra-cell current reported in Eq.(\ref{Jjintra-def}). 

\subsection{Drude weight}\label{App-Drude}
The Drude weight $D$ characterizing the LRT of the SSH model can be  computed  following  Kohn's approach[\onlinecite{kohn_1964}]. In particular, in a 1D ring one has 
\begin{equation}
D=-L \frac{d^2 E_0(\Phi)}{d^2 \Phi}\bigg|_{\Phi=0} \nonumber
\end{equation}
where $E_0(\Phi)$ denotes the dependence of the many-body ground state energy on the magnetic flux $\Phi$ threading the ring, while $L$ denotes the ring  length. For a tight-binding model, we can associate the magnetic flux $\Phi$ to a phase $\phi$ in the hopping amplitudes according to $N\phi=2\pi \Phi/\Phi_0$, where $N$ is the number of links in the ring and $\Phi_0=h/e$ is the magnetic flux quantum. Hence, in a bipartite lattice with two sites per cell, we get:
\begin{equation}\label{Phi-phi}
\Phi=L \, \frac{2}{a} \, \frac{\hbar}{e} \, \phi 
\end{equation}
where $a$ is the lattice constant. Exploiting the linear relation between $\phi$ and $\Phi$ we can write Kohn's formula as
\begin{equation}\label{Drude-Kohn}
D=- \left(\frac{a}{2}\right)^2 \left(\frac{e}{\hbar}\right)^2 L^{-1} \frac{d^2 E_0(\phi)}{d^2 \phi}\bigg|_{\phi=0}   
\end{equation}
Moreover, for translation invariant one-body Hamiltonians we can write:
\begin{eqnarray}
D&=&-\left(\frac{a}{2}\right)^2 \left(\frac{e}{\hbar}\right)^2  L^{-1} \left[ \frac{d^2 }{d^2 \phi} \sum_{(k,b)\in \mathcal{I}} \varepsilon_b(k,\phi) \right]_{\phi=0} \nonumber 
\end{eqnarray}
where $\mathcal{I}$ denotes the set of bands $b$ and wavevectors $k$ that are occupied in the many body ground state without flux, while $\varepsilon_b(k,\phi)$ denotes the band dispersion relations for a finite flux. 

Since the single particle energies depend on the phase $\phi$ only through the combination $ka+2\phi$, see Eq.(\ref{SSH-spectrum}) in the Main Text, the many-body ground state energy of a half filled SSH model with dimerization ($r \ne 1$) does not depend on the flux in the thermodynamic limit. Indeed, due to the periodic nature of the lower band over the interval $ka\in[-\pi,\pi]$, one has
\begin{eqnarray}
L^{-1} E_0^{r\ne 1}(\phi)&=&\frac{1}{2\pi} \frac{1}{a} \int_{-\pi}^\pi d (ka) \varepsilon_-(k,\phi) \nonumber \\
&=&\frac{1}{2\pi} \frac{1}{a} \int_{-\pi}^\pi d (ka) \varepsilon_-(k,0)= L^{-1} E_0^{r\ne 1}(0) \nonumber
\end{eqnarray}
and we conclude that the Drude weight  (\ref{Drude-Kohn}) is identically zero in this case, consistently with Eq.(\ref{Jdc-insulating}) of the Main Text.

Without dimerization ($r=1$) the situation is different, since  we get
\begin{eqnarray}
L^{-1} E_0^{r= 1}(\phi)
&=&-\frac{v}{\pi} \frac{2}{a} \int_{-\frac{\pi}{2}}^{\frac{\pi}{2}} d \big(\frac{ka}{2} \big)  \cos\big(\frac{ka}{2} + \phi \big) \nonumber \\
&=&-\frac{2v}{\pi} \frac{2}{a} \cos( \phi ) \nonumber
\end{eqnarray}
and from Eq.(\ref{Drude-Kohn}) one obtains
\begin{equation}
D=-\frac{e^2}{\hbar} \frac{2v}{\hbar \pi} \frac{a}{2} = -\frac{e^2}{\hbar} \frac{\mathtt{v}_F}{ \pi} \nonumber
\end{equation}
where $\mathtt{v}_F$ is the Fermi velocity. The finite Drude weight computed in this way coincides with the one obtained in the Main Text (see Eq.(\ref{Jdc-metallic})) through its dynamical definition.



\begin{thebibliography}{99}
\bibitem{footnote_Drude} The expression for $D$ is obtained by re-expressing the phase $\phi_i$ in terms of the the initial magnetic flux $\Phi_i=\mathcal{E}_i L$ (see Eq.(\ref{Phi-phi}) in Appendix \ref{App-Drude}). Moreover an additional factor $e$ has to be included to obtain the charge current from the particle current $J$ in Eq.(\ref{Jdc-metallic}).
\end{thebibliography}
 \end{document}